\documentclass[12pt,a4paper]{article}

\newcommand{\pa}{\partial}
\newcommand{\ma}{\mathcal{A}}
\newcommand{\sn}{\mathrm{sn}}
\newcommand{\cn}{\mathrm{cn}}
\newcommand{\qq}{\qquad}

\begin{document}

\begin{flushright}
hep-th/0209218
\end{flushright}
\vspace{1.8cm}

\begin{center}
 \textbf{\Large Penrose Limits of Branes  \\ 
and Marginal Intersecting Branes}
\end{center}
\vspace{1.6cm}
\begin{center}
 Shijong Ryang
\end{center}

\begin{center}
\textit{Department of Physics \\ Kyoto Prefectural University of Medicine
\\ Taishogun, Kyoto 603-8334 Japan}
\par
\texttt{ryang@koto.kpu-m.ac.jp}
\end{center}
\vspace{2.8cm}
\begin{abstract}
We construct the Penrose limit backgrounds in closed forms along the 
generic null geodesics for the near-horizon geometries of D1, D3, D5,
NS1 and NS5 branes. The Penrose limit metrics of D1, D5 and NS1 have
non-trivial dependence of the light-cone time coordinate, while those
of D3 and NS5 have no its dependence. We study the Penrose limits
on the marginal 1/4 supersymmetric configurations of standard 
intersecting branes, such as the NS-NS intersection of NS1 and NS5,
the R-R intersections of D$p$ and D$q$ over some spatial dimensions
and the mix intersections of NS5 and D$p$ over ($p$ -1)-dimensional
spaces. They are classified into three types that correspond to
the Penrose limits of D1, D3 and D5 backgrounds.
\end{abstract}
\vspace{3cm}
\begin{flushleft}
September, 2002 \\
\end{flushleft}

\newpage
\section{Introduction}

The Penrose limit \cite{PE,RG} on the $AdS_5 \times S^5$ solution of 
type IIB theory has been shown to yield an interesting pp-wave background
with maximal supersymmetry \cite{BFH,BFP}. The string theory in
this background is exactly solvable for the spectrum of oscillators
\cite{RM,MT} so that we can explicitly observe the duality in a 
particular sector of the four-dimensional $\mathcal{N} = 4$ SYM gauge
theory with the type IIB string theory in the pp-wave background
beyond the supergravity approximation \cite{BMN}.
Moreover, the D-branes on the pp-wave background 
have been explored \cite{DP}

According to the three kinds of classifications of null geodesics such as
longitudinal, radial and generic ones for a supergravity brane solution,
there are three corresponding Penrose limits \cite{BFP}. The Penrose
limit along the longitudinal null geodesic yields the trivial flat
Minkowski spacetime, while the limit along the radial one is 
performed to construct the Penrose limit metrics for the backgrounds
such as  D$p$-brane,  fundamental string,  NS5-brane, M-brane and 
so on. The near-horizon geometries of D3-brane, M2-brane
and M5-brane have themselves AdS structure and turn out to be
the pp-wave backgrounds in the Penrose limit along the generic
null geodesic. Both M2-brane and M5-brane configurations have the 
same Penrose limit metric, where there is an isometric symmetry
between two configurations. 
The standard intersecting systems such as three M2-branes transversely
intersecting over a 0-brane, two M2-branes and two M5-branes transversely
intersecting over a 0-brane have been investigated and the Penrose limit
metrics along the generic null geodesics on their near-horizon geometries
that include AdS structure have 
been shown to include the metric of the
Cahen-Wallach space. The Penrose limits along the generic null geodesics
have been studied for the supersymmetric black holes in four and 
five dimensions and the supersymmetric string in six dimensions, 
whose near-horizon geometries also include AdS structure, and
shown to consist of the Cahen-Wallach space.  

The other pp-wave background has been found by taking the Penrose limit
on the near-horizon geometry of
a standard intersecting system that two D3-branes transversely 
intersect over a string,  which is also expressed by 
$AdS_3\times S^3\times T^4$ \cite{CLP}. Another pp-wave solution
has been constructed by performing the Penrose limit along the
generic null geodesic on a non-standard intersection of two NS5-branes
over a string,  where the harmonic function for each brane component
depends on the coordinates of the relative transverse space rather those
of the overall transverse space \cite{LP,KNS}.
Various types of non-standard intersecting systems whose near-horizon
geometries are product spaces including $AdS_p\times S^q$ have been also
demonstrated to yield the pp-wave solutions in the Penrose limit
\cite{LP}.

The Penrose limit on the near-horizon geometry of NS5-brane has been 
taken to yield the pp-wave background that is dual to a high energy
sector of the little string theory \cite{HRV}. For the geometries of 
D$p$-branes and their near-horizon limits the Penrose limits along
the generic null geodesic have been studied \cite{GZS,FIS}, where
the Penrose limits of metric, dilaton and ($p+2$)-form field strength
are expressed  in terms of the radial  coordinate 
that is related implicitly with a light-cone time coordinate. 

Following the framework of Ref. \cite{BFP} we will try to construct 
the Penrose limit backgrounds explicitly in terms of the light-cone
time coordinate itself  for the near-horizon geometries
of D$p$-branes with $p = 1, 3, 5.$  Similarly the Penrose limits of
the near-horizon geometries of NS1-brane and NS5-brane will be derived
in closed forms and compared with those of D$p$-branes. Further
we will demonstrate the various Penrose limits of the marginal BPS
configurations expressed by the standard intersecting branes such as
the intersection of an NS1-brane and an NS5-brane over a string,
that of a D$p$-brane and a D$q$-brane over a $n$-brane with
$n = ( p + q )/2 - 2, n = 0, 1, 2$ and that of an NS5-brane 
and a D$p$-brane over a $(p-1)$-brane. 
Comparing the Penrose limits of these various
intersecting branes with those of D$p$-branes with $p = 1, 3, 5.$
we will classify the Penrose limits of these marginal BPS bound 
states into three types.

\section{Penrose limit along the generic null geodesic}

We start to review the relevant aspects of the Penrose limit in 
Ref. \cite{BFP}. We consider the Penrose limit of a ten-dimensional 
background with a metric
\begin{equation}
ds^2 = A^2(-dt^2 + ds^2(E^p)) + B^2 dr^2 + B^2 r^2(d\psi^2 + 
\sin^2 \psi d\Omega_{7-p}^2),
\label{met}\end{equation}
where the metric on $S^{8-p}$ has been written out as   
$d\Omega_{8-p}^2 = d\psi^2 + \sin^2 \psi 
d\Omega_{7-p}^2$. We choose a null
geodesic to lie in the $(t, r, \psi)$ plane and perform the coordinate 
transformation from $(t, r, \psi)$ to $(u, v, \tilde{z})$
\begin{equation}
u = u(r), \qq v = t + l\psi + a(r), \qq \tilde{z} = \psi + b(r),
\end{equation}
where $u$ is the affine parameter along the generic null
geodesic and $l$ is constant. If $a(r), u(r)$ and $b(r)$ are specified by
$da/dr = (B^2/A^2 - l^2/r^2)^{1/2}$,
\begin{equation}
u = \int^r \frac{B^2 dr}{\sqrt{\frac{B^2}{A^2} - \frac{l^2}{r^2}}}  
\label{ur}\end{equation} 
and
\begin{equation}
b = - \int^r \frac{l/r^2\; dr}{\sqrt{\frac{B^2}{A^2} - \frac{l^2}{r^2}}}
\label{br}\end{equation}
and the Penrose limit is taken along this null geodesic after the 
rescaling of coordinates, then the metric is expressed in terms
of Rosen coordinates as 
\begin{equation}
ds^2 = 2dudv + (B^2r^2 - l^2A^2)d\tilde{z}^2 + A^2\sum_{a=1}^{p}
(d\tilde{x}^a)^2 + B^2r^2\sin^2b\sum_{i=1}^{7-p}(d\tilde{y}^i)^2.
\label{mro}\end{equation}
This is the metric of spacetime in the neighbourhood of the generic
null geodesic in the specific Penrose scaling limit. For the D$p$-brane
solution characterized by a harmonic function $H = C^{-1} 
= 1 + Q_p/r^{7-p}$ with the D$p$-brane charge $Q_p$, the R-R 
$(p+2)$-form field strength $F_{p+2} = dvol(E^{1,p})\wedge dC(r)$
becomes in the Penrose limit to be
\begin{equation}
\bar{F}_{p+2} = C'\frac{l}{B}\sqrt{\frac{1}{A^2} - \frac{l^2}{B^2r^2}}
du\wedge d\tilde{x}^1 \wedge \cdots d\tilde{x}^p \wedge d\tilde{z},
\label{fst}\end{equation}
where $A^{-2} = B^2 = H^{1/2}$.
Using the following change of coordinates 
\begin{eqnarray}
u &=& x^-, \nonumber \\
v &=& x^+ + \frac{\pa_-A}{2A} x^2 + \frac{\pa_-(rB\sin b)}{2rB\sin b} y^2
+ \frac{\pa_- \sqrt{B^2r^2 - l^2A^2}}{2\sqrt{B^2r^2 - l^2A^2}}z^2, 
\label{coc} \\
\tilde{x}^a &=& \frac{x^a}{A},\qq \tilde{y}^i = \frac{y^i}{rB\sin b},\qq
\tilde{z} = \frac{z}{\sqrt{B^2r^2 - l^2A^2}}, \nonumber
\end{eqnarray}
where $\pa_- = d/dx^-, A = A(r(x^-)), B = B(r(x^-)), r = r(x^-)$ that is
implicitly given by inverting the Eq. (\ref{ur}), we can transform 
the metric (\ref{mro}) in the Rosen form into in the Brinkman form
\begin{equation}
ds^2 = 2dx^+dx^- + \ma (dx^-)^2 + \sum_{a=1}^{p}(dx^a)^2
+ \sum_{i=1}^{7-p}(dy^i)^2 + dz^2,
\end{equation}
where
\begin{equation}
\ma = \frac{\pa_-^2A}{A} x^2 + \frac{\pa_-^2(rB\sin b)}{rB\sin b} y^2
+ \frac{\pa_-^2 \sqrt{B^2r^2 - l^2A^2}}{\sqrt{B^2r^2 - l^2A^2}}z^2.
\label{afa}\end{equation}
 
\section{Penrose limits of elementary branes}

We are ready to consider the near-horizon geometry of D3-brane 
configuration characterized by $A^{-2} = B^2 = \sqrt{Q_3}/r^2$ for the
metric (\ref{met}). The original radial coordinate is explicitly 
determined from (\ref{ur}) as
\begin{equation}
r = \frac{\sqrt{Q_3}}{l} \sin \left( \frac{l}{\sqrt{Q_3}} u \right),
\label{rsi}\end{equation}
where we have simply chosen an integration constant in such a way as
$u$ becomes zero at $r = 0$. This solution yields a restriction
$Q_3/l^2 \ge r^2$. Since $B^2r^2 - l^2A^2 = \sqrt{Q_3}
\cos^2(lu/ \sqrt{Q_3})$ and $b =  -lu/ \sqrt{Q_3}$ where $b$ in 
(\ref{br}) is also chosen to be zero at $r=0$, the non-trivial
factor $\ma$ (\ref{afa}) in the metric is calculated by
\begin{equation}
\ma = -\frac{l^2}{Q_3}\left(\sum_{a=1}^3(x^a)^2 + 
\sum_{i=1}^4(dy^i)^2 + z^2 \right),
\end{equation}
which is negative definite. Even if we take account of the integration
constants as $r = (\sqrt{Q_3}/l)\sin (lu/\sqrt{Q_3} + C_1), 
b = -lu/\sqrt{Q_3} + C_2$, we get the same result. Applying the 
coordinate transformation (\ref{coc}) to the R-R field strength 
(\ref{fst}) for the near-horizon geometry of D$p$-brane with 
$H = Q_p/r^{7-p}$ we have 
\begin{eqnarray}
\bar{F}_{p+2} &=& C'\frac{l}{rB^2A^{p+1}}
dx^-\wedge dx^1 \wedge \cdots dx^p \wedge dz \nonumber \\
 &=& (7-p)lQ_p^{\frac{p-5}{4}}r^{\frac{(9-p)(3-p)}{4}}dx^-\wedge dx^1 
\wedge \cdots dx^p \wedge dz.
\label{fpr}\end{eqnarray}
The dilaton is simply given by
\begin{equation}
e^{2\phi} = \left( \frac{Q_p}{r^{7-p}} \right)^{\frac{3-p}{2}}.
\label{phr}\end{equation}
The near-horizon geometry of D3-brane is special since the coefficient in
(\ref{fpr}) becomes a constant $4l/\sqrt{Q_3}$, where we have to add  
the Hodge dual for the $p = 3$ case. This  background gives the solution
of the maximally supersymmetric pp-wave in the type IIB theory.
When $l$ is taken to be zero, both the 5-form field strength and the
factor $\ma$ vanish. Consequently the pp-wave background reduces to the 
ten-dimensional Minkowski spacetime that is the Penrose limit on the
near-horizon geometry of D3-brane along the radial null geodesic.

Now we consider the near-horizon limit of the D5-brane metric, that is
(\ref{met}) with $A^2 = B^{-2} = r/\sqrt{Q_5}$. From (\ref{ur}) $r$ is
expressed in terms of $u$ as 
\begin{equation}
r = \sqrt{\frac{Q_5 - l^2}{Q_5}}u,
\label{ru}\end{equation}
where an integration constant is chosen such that $u = 0$ at $r = 0$, and
$l$ is bounded as $l^2 < Q_5$. The integration in (\ref{br}) leads to
\begin{equation}
b = - \frac{l}{\sqrt{Q_5 - l^2}} \ln \frac{r}{r_1}
\label{bln}\end{equation}
with an integration constant $r_1$. Substituting (\ref{ru}) and 
(\ref{bln}) into (\ref{afa}) we can derive
\begin{equation}
\ma = - \frac{1}{4(x^-)^2}\left[ \sum_{a=1}^5 (x^a)^2 + z^2 + 
 \left( 1 + \frac{4l^2}{Q_5 - l^2} \right)
\sum_{i=1}^2 (y^i)^2 \right],
\label{afi}\end{equation}
which is negative definite due to $Q_5 > l^2$. It is noted that the result
is independent of the integration constant $r_1$ for (\ref{br}).
If we choose the other integration constant 
for $u=u(r)$ (\ref{ur}), the resulting expression of
$\ma$ simply shows the shift of $x^-$. Therefore any integration conatant
for (\ref{ur}) can be absorbed into the 
definition of the light-cone time coordinate.
From (\ref{fpr}) the R-R 7-form field strength 
is specified by the non-constant value
\begin{equation}
\frac{2lQ_5}{Q_5 - l^2}\frac{1}{(x^-)^2},
\label{lqx}\end{equation} 
while the dilaton in (\ref{phr}) is also a function of $x^-$
\begin{equation}
e^{2\phi} = \frac{Q_5 - l^2}{Q_5^2} (x^-)^2.
\label{pfi}\end{equation}

For the near-horizon geometry of D1-brane specified by $A^2 = B^{-2} =
r^3/\sqrt{Q_1}$, we have to manipulate the integration
\begin{equation}
u = \int_0^r dr \frac{1}{\sqrt{1 - \frac{l^2}{Q_1} r^4}},
\label{ufi}\end{equation}
where an integration constant is fixed as $u=0$ at $r =0$. We restrict
the region of $r$ to $Q_1/l^2 \ge r^4$ where the null geodesic line is
well defined, and then the Eq. (\ref{ufi}) can be described as
\begin{equation}
- \frac{\sqrt{2l}}{Q_1^{1/4}}u + K\left(\frac{1}{\sqrt{2}}\right) =
F\left( \cos^{-1} \left(\frac{\sqrt{l}}{Q_1^{1/4}} r \right), 
\frac{1}{\sqrt{2}} \right),
\end{equation}
where $K(1/\sqrt{2})$ is complete elliptic integral of the first kind and
$F(\phi, 1/\sqrt{2})$ is incomplete one. It is possible to invert this 
equation as
\begin{equation}
\sqrt{1- \frac{l}{\sqrt{Q_1}} r^2} = \sn \left( w, 
\frac{1}{\sqrt{2}}\right),
\label{sn}\end{equation}
where $w = -\sqrt{2l} u/Q_1^{1/4} + K$ and $\sn z$ is one of Jacobi's 
elliptic functions with a property $\sn K =1$, that yields $u=0$ at $r=0$.
The original radial coordinate $r$ is explicitely described in
terms of  $u$ as
\begin{equation}
r = \frac{Q_1^{1/4}}{\sqrt{l}} \cn w,
\label{rcn}\end{equation}
which indeed satisfies $Q_1/l^2 \ge r^4$ and correctly reduces to $r=u$
in the $l\rightarrow 0$ limit. The radial 
coordinate $r$ is the periodic function
of $u$, which is similar to the D3-brane case.  
Choosing an integration constant in (\ref{br}) such
that $b$ vanishes at $r=0$, we have $\sin 2b =-lr^2/\sqrt{Q_1}$
which yields
\begin{equation}
\sin b = - \frac{1}{\sqrt{2}} \left( 1- \sqrt{1- \frac{l^2}{Q_1}r^4}
\right)^{1/2}.
\label{sib}\end{equation}
Using this expression and $dr/du = \sqrt{2}\sn w \mathrm{dn} w$
that resuces to $(1-l^2r^4/Q_1)^{1/2}$ 
through (\ref{sn}), which is also given from
(\ref{ufi}), we compute the factor $\ma$ as a function of $r$
\begin{equation}
\ma = -\frac{3}{4r^2Q_1} [ (5l^2r^4 - Q_1)(x_1^2 + z^2) +
(l^2r^4 - Q_1)\sum_{i=1}^6 (y^i)^2 ],
\label{ari}\end{equation}
which is further expressed in terms of $x^-$ as
\begin{equation}
\ma = - \frac{3l}{4\sqrt{Q_1}\cn^2w}[(5\cn^4w - 1)(x_1^2 + z^2) +
(\cn^4w - 1)\sum_{i=1}^6 (y^i)^2 ]
\label{air}\end{equation}
with $w = -\sqrt{2l}x^-/Q_1^{1/4} + K$. The factor $\ma$ is singular at
$x^- =0$ due to $\cn K=0$. The R-R 3-form field strength and the dilaton
in the Penrose limit are given by
\begin{equation}
\bar{F}_{x^-x^1z} = \frac{6}{l} \cn^4w, \qq e^{2\phi} = 
\frac{l^3}{\sqrt{Q_1}}\frac{1}{\cn^6w},
\label{fph}\end{equation}
which show the non-trivial $x^-$ dependences. 

Here we describe the general expression of the Penrose limit metric for
the near-horizon geometry of Dp-brane in terms of the original radial
coordinate. The non-trivial component of the metric (\ref{afa}) is 
evaluated as 
\begin{eqnarray}
\ma &=& - \frac{7-p}{16r^2}\Biggl[ \left( (p-3) + \frac{(13-3p)l^2}{Q_p}
r^{5-p} \right)(\sum_{a=1}^p (x^a)^2 + z^2 ) \nonumber \\
 &+& \left( (p-3) +
\frac{(p+1)l^2}{Q_p} r^{5-p} \right) \sum_{i=1}^{7-p} (y^i)^2 \Biggr],
\label{adp}\end{eqnarray}
where we have used $du/dr = 1/\sqrt{1-l^2r^{5-p}/Q_p}, db/dr = 
-l /\sqrt{Q_pr^{p-3} - l^2r^2}$ and we have observed that the $\cos b$
terms in $\pa_u^2(rB\sin b)$ cancel out. In Ref. \cite{GZS} the same
implicit expression of the factor $\ma$ for the D$p$-brane was presented
and expressed in terms of a variable that is defined to be proportional
to $r^{(p-5)/2}$ for $p \neq 5$. Recently the same expression described
in terms of $r$ itself has been presented in Ref. \cite{FIS}.
For $p=3$ and $p=5$ the factor $\ma$ in (\ref{adp}) turns out to be
fairely simplified. For $p=1$ it reduces to (\ref{ari}).
In the $l \rightarrow 0$ limit $u$ is equal to $r$, that is seen in
(\ref{rsi}), (\ref{ru}), (\ref{rcn}), and the factor $\ma$
becomes a symmetric expression
\begin{equation}
\ma = - \frac{(7-p)(p-3)}{16(x^-)^2}(\sum_{a=1}^p (x^a)^2 + z^2
+ \sum_{i=1}^{7-p} (y^i)^2),
\end{equation}
which gives the Penrose limit metric of the near-horizon geometry of
D$p$-brane along the radial null geodesic.
Thus we have performed the explicit integration in $u=u(r)$ and
derived its inversion $r=r(u)$ so that the factor $\ma$ 
together with the dilaton and the R-R field strength has been expressed
in terms of the Brinkman coordinate $x^-$ for the D1, D3 and D5 
configurations. It is noted that there is a common structure between 
(\ref{afi}) the factor $\ma$ of D5-brane and (\ref{air}) the factor $\ma$
of D1-brane that each $\ma$ is singular at $x^- = 0$, which is obliged
to $1/r^2$ factor in (\ref{adp}). On the other hand there is a reciprocal
structure that $\bar{F}_3$ and $e^{2\phi}$ in (\ref{fph}) 
for the D1-brane vanishes and diverges at $x^- = 0$ respectively,
while $\bar{F}_7$ in (\ref{lqx}) and $e^{2\phi}$ in (\ref{pfi}) for
the D5-brane diverges and vanishes respectively. This reciprocal
behavior of the $x^-$ dependence can be read from (\ref{fpr}) and 
(\ref{phr}), where $p = 3$ is the critical number.

Now we turn our attention to the near-horizon geometry of NS5-brane.
This geometry is so specified by $A =1, B = Q_5^{NS}/r^2$ with
the charge of NS5-brane $Q_5^{NS}$ in (\ref{met}) that we have
\begin{equation}
r = C_1 e^{\frac{\sqrt{Q_5^{NS}-l^2}}{Q_5^{NS}} u}, \qq 
b = -\frac{l}{Q_5^{NS}}(u + C_2)
\end{equation}
with the integration constants $C_1, C_2$. The $u$ dependence of $b$ is
the same as the D3-brane case. Combining these expressions leads to the
metric in the Penrose limit characterized by
\begin{equation}
\ma = - \frac{l^2}{(Q_5^{NS})^2} \sum_{i=1}^2 (y^i)^2,
\end{equation}
where there are no $x^2$ and $z^2$ terms. This result agrees
with the expression presented in Ref. \cite{HRV,FIS}.
The $x^-$ dependence as well as the $C_1$ and $C_2$ dependences do not
appear in the same way as the previous D3-brane case, where
$rB = contstant$ in both cases. 
   
For the near-horizon geometry of the fundamental string specified by
$B =1, A^2 = H^{-1}, H = Q_1^{NS}/r^6$ with 
the charge of NS1-brane $Q_1^{NS}$,
we can describe $r$ explicitly in terms of $u$ by performing the 
integration in (\ref{ur}) as
\begin{equation}
u = \frac{\sqrt{Q_1^{NS}} - \sqrt{Q_1^{NS} - l^2r^4}}{2l^2}
\label{uqr}\end{equation}
and inverting it
\begin{equation}
r = \frac{1}{\sqrt{l}}[Q_1^{NS} - (2l^2 u - \sqrt{Q_1^{NS}})^2
]^{1/4},
\label{rns}\end{equation}
which vanishes at $u=0$. The null geodesic line is well defined for
$Q_1^{NS}/l^2 \ge r^4 \ge 0$.  Through (\ref{uqr}) 
this region is mapped to
$0 \le u \le \sqrt{Q_1^{NS}}/2l^2$. On the other hand by carrying out
the integration in (\ref{br}) we have the same expression as the
D1-brane case
\begin{equation}
\sin 2b = - \frac{lr^2}{\sqrt{Q_1^{NS}}},
\end{equation}
where an integration constant is also chosen as $b=0$  at $r=0$.
Similarly to the D1-brane case we can extract $\sin b$ in the same 
expression as (\ref{sib}). The substitution of (\ref{rns}) into
(\ref{sib}) with $Q_1^{NS}$
yields $\sin b = -l\sqrt{u}/(Q_1^{NS})^{1/4}$. Gathering together we 
derive the factor $\ma$ in the Penrose limit
\begin{equation}
\ma = - \frac{3l^4}{(Q_1^{NS} - (2l^2 x^- - \sqrt{Q_1^{NS}})^2)^2}
[(2Q_1^{NS} - (2l^2 x^- - \sqrt{Q_1^{NS}})^2)(x_1^2 + z^2) 
+ Q_1^{NS}\sum_{i=1}^6 (y^i)^2]
\label{ans}\end{equation}
for $0 \le x^- \le \sqrt{Q_1^{NS}}/2l^2$, which again shows the
coincidence of coefficients of the $x_1^2$ 
and $z^2$ terms and the singular
behavior at $x^- = 0$. Here we can rewrite (\ref{ans}) in terms of 
$r = (Q_1^{NS} - (2l^2 x^- - \sqrt{Q_1^{NS}})^2)^{1/4}/\sqrt{l}$ as
\begin{equation}
\ma = - \frac{3}{r^8} [ ( Q_1^{NS} + l^2r^4 )(x_1^2 + z^2) + Q_1^{NS}
\vec{y}^2],
\label{anr}\end{equation}
whose $1/r^8$ behavior is compared with $1/r^2$ in (\ref{adp})
for the D$p$-branes. This expression in terms of $r$ agrees with
the result in Ref. \cite{FIS}.

 The Penrose limit for the NS-NS 3-form field 
strength $F_3^{NS}=dt\wedge d\tilde{x_1} \wedge dH^{-1}$ and the
dilaton leads to
\begin{eqnarray}
\bar{F}_{x^-x^1z}^{NS} &=& \frac{6l^2}{[Q_1^{NS} - 
(2l^2 x^- - \sqrt{Q_1^{NS}})^2]^{1/2}}, \nonumber \\
e^{-2\phi} &=& \frac{Q_1^{NS} l^3}
{[Q_1^{NS} - (2l^2 x^- - \sqrt{Q_1^{NS}})^2]^{3/2}}.
\end{eqnarray}
When $l$ is taken to be zero, we have
\begin{equation}
\ma = - \frac{3}{16(x^-)^2} ( x_1^2 + z^2 + \vec{y}^2),\qq
e^{-2\phi} = \frac{(Q_1^{NS})^{1/4}}{8}\frac{1}{(x^-)^{3/2}},\qq
\bar{F}_{x^-x^1z}^{NS} = 0,
\end{equation}
which reproduce the Penrose limit of the near-horizon geometry of
NS1-brane along the radial null geodesic \cite{BFP}.
   
\section{Penrose limits of intersecting branes}

Let us consider the Penrose limits on the near-horizon geometries of
intersecting brane configurations along the generic null geodesics
whose tangent vectors have a component tangent to the  overall transverse
sphere. There are the following standard intersections representing
the marginal 1/4 supersymmetric bound states :
(a) NS-NS intersections: NS1$\parallel$NS5 ( the internal dimensions of 
an NS1-brane and an NS5-brane are parallel); (b) R-R intersections:
D$p\perp$D$q\; (n), n = (p+q)/2 -2$ (a D$p$-brane overlapps a 
D$q$-brane in a $n$-dimensional space ); (c) mixed intersections:
NS5$\perp$D$p\; (n), n = p-1$ \cite{AT,PT}.

We first consider the near-horizon limit of a fundamental string 
smeared over a solitonic NS5-brane with metric
\begin{equation}
ds^2 = H_1^{-1}(-dt^2 + dx_1^2) + \sum_{a=2}^5 (dx^a)^2 +
H_5^{NS}[dr^2 + r^2(d\psi^2 + \sin^2\psi d\Omega_2^2)]
\end{equation}
with $H_1^{NS} = Q_1^{NS}/r^2, H_5^{NS} = Q_5^{NS}/r^2.$ 
Since the $u$ in (\ref{ur}) and the $b$ in (\ref{br}) show the same 
behaviors as the D3-brane case, we derive  the Penrose limit metric
through the appropriate coordinate transformations
\begin{equation}
ds^2 = 2dx^+ dx^- + \ma (dx^-)^2 + \sum_{a=1}^5 (dx^a)^2 + 
\sum_{i=1}^2(dy^i)^2 + dz^2,
\end{equation}
where 
\begin{equation}
\ma = -\frac{l^2}{(Q_5^{NS})^2}(x_1^2 + z^2 + \sum_{i=1}^2 (y^i)^2)
\label{mnn}\end{equation}
and the $Q_1^{NS}$ dependence does not emerge. It is observed that this
$x^-$-independent metric describes a lorentzian symmetric or 
Cahen-Wallach space. The constant $l^2/(Q_5^{NS})^2$ can be absorbed
into a boost of $(x^+ , x^-)$. 

For the R-R intersections with $n=1$, D5$\parallel$D1, D4$\perp$D2(1),
D3$\perp$D3(1), we see that $u$ and $b$ are also identical to those
of the D3-brane case and evaluate the factor $\ma$ of th Penrose limit 
metric to be the same form as (\ref{mnn})
\begin{equation}
\ma = - \frac{l^2}{k} (x_1^2 + z^2 + \sum_{i=1}^2(y^i)^2),
\label{arr}\end{equation}
where $k = Q_5Q_1, Q_4Q_2$ and $Q_3Q_3'$ respectively, and $x_1$ is the 
coordinate of the common spatial direction of two intersecting D-branes.
In the starting metric the coefficients of $\sum_{a=2}^5(dx^a)^2$  for the
internal coordinates of the D$p$-brane and the D$q$-brane 
are constant and characterized by the ratio $Q_p/Q_q$ so that
there are no $x_a^2$ terms for $a = 2, \cdots 5$ in $\ma$. 
The result (\ref{arr}) for the D3 $\perp$ D3 (1) configuration
is in agreement with the Penrose limit metric studied in Ref. \cite{CLP}.
The near-horizon geometries of these R-R intersections are represented by
$AdS_3 \times S^3 \times T^4$ in the same way as the near-horizon geometry
of the NS1$\parallel$NS5 intersection, 
whose $AdS_3 \times S^3$ part becomes
the six-dimensional Cahen-Wallach space in the Penrose limit
along the generic null geodesic.  
For the other $n=0$ case, D4$\parallel$D0, D3$\perp$D1, D2$\perp$D2
the integrations in $u(r)$ and $b(r)$ are the same as those for
D1-brane case and then the factor $\ma$ can be expressed as
\begin{equation}
\ma = - \frac{3l}{4k\cn^2w}[(5\cn^4w - 1)z^2 +
(\cn^4w - 1)\sum_{i=1}^3 (y^i)^2 ]
\end{equation}
with $w = -\sqrt{2l}x^-/\sqrt{k} + K$, where $k= \sqrt{Q_4Q_0},
\sqrt{Q_3Q_1}$ and $\sqrt{Q_2Q_2'}$ respectively.
There is another $n=2$ case where D6$\parallel$D2, D5$\perp$D3(2) and
D4$\perp$D4(2) configurations are the threshold BPS bound states.
In this case the overall transverse space is three-dimensional so that
$u$ is linear to $r$ and $b$ is a logarithmic function of $r$ in the
same form as the D5-brane case (\ref{ru}), (\ref{bln}). 
The factor $\ma$ in the Penrose limit metric is similarly 
provided by
\begin{equation}
\ma = - \frac{1}{4(x^-)^2}\left[\sum_{i=1}^2 (x^i)^2 + z^2 +
\left( 1 + \frac{4l^2}{k - l^2} \right) y_1^2 \right],
\end{equation}
where $k = Q_6Q_2, Q_5Q_3$ and $Q_4Q_4'$ respectively and the
D$p$-brane and the D$q$-brane orthgonally overlap in the $x^1$ and $x^2$
directions. In the $n=3$ case the overall transverse space of 
D5$\perp$D5(3), D4$\perp$D6(3) is two-dimensional. Since the relevant
harmonic function is expressed by a logarithmic function, we cannot
explicitly perform the integration in 
$u(r), b(r)$ and then it is impossible to
obtain the Penrose limit metric in a closed form. 

The Penrose limits of the R-R field strength and the dilaton 
will be demonstrated for two examples, D2$\perp$D2 and D4$\perp$D4(2).
For the D2$\perp$D2 configuration the R-R 4-form 
\begin{equation}
F_4 = dt\wedge(dx^1\wedge dx^2\wedge dH_2^{-1} + dx^3\wedge dx^4\wedge
d{H'}_2^{-1})
\end{equation}
 is scaled in the Penrose limit to be 
\begin{equation}
\bar{F}_4 = \frac{3\cn^{5/2}w}{l^{1/4}(Q_2Q'_2)^{1/8}} dx^-\wedge
(dx^1\wedge dx^2 + dx^3\wedge dx^4)\wedge dz
\end{equation}
by taking account of the appropriate constant transformations of
$x^i (i=1, \cdots 4)$ coordinates.
The dilaton specified by $e^{-2\phi}=(H_2H'_2)^{-1/2}=r^3/\sqrt{Q_2Q'_2}$
turns out to be a $x^-$-dependent function 
$e^{-2\phi} =(Q_2Q'_2)^{1/4}\cn^3w/l^{3/2}$,
which shows some similar behavior to the D-1brane case, that is,
the square root of (\ref{fph}) accompanied with $\phi \rightarrow 2\phi,
Q_1 \rightarrow Q_2Q'_2$. Similarly 
we consider the near-horizon geometry
of D4$\perp$D4(2). The R-R 6-form field strength is represented by
\begin{equation}
F_6 = dt\wedge dx^1\wedge dx^2\wedge (dx^3\wedge dx^4\wedge dH_4^{-1}
+ dx^5\wedge dx^6\wedge d{H'}_4^{-1}),
\end{equation}
where one intersecting D4-brane extends to $(x^1, x^2, x^3, x^4)$ 
directions and the other intersecting D4-brane to
$(x^1, x^2, x^5, x^6)$ directions. In the Penrose limit it is given by
the $x^-$-dependent expression
\begin{equation}
\bar{F}_6 = \frac{l\sqrt{Q_4Q'_4}}{(Q_4Q'_4 - l^2)^{3/4}(x^-)^{3/2}}
dx^- \wedge dx^1\wedge dx^2\wedge (dx^3\wedge dx^4 +  dx^5\wedge dx^6)
\wedge dz.
\end{equation}
The dilaton described by $e^{-2\phi} = (H_4H'_4)^{1/2}$ 
becomes in the Penrose limit to be
specified by $e^{-2\phi} = Q_4Q'_4/(x^-\sqrt{Q_4Q'_4 - l^2})$
 which is similar to the D5-brane case, that is,
the square root of (\ref{pfi}) accompanied with $\phi \rightarrow 2\phi,
Q_5 \rightarrow Q_4Q'_4$.

Now let us consider the remaining mix intersection, NS5$\perp$D$p(p-1)$
for $p=1, \cdots 6$ with a metric 
\begin{equation}
ds^2 = H_p^{-1/2}[-dt^2 + \sum_{i=1}^{p-1}(dx^i)^2 + H_5^{NS}(dx^p)^2]
+ H_p^{1/2}[\sum_{i=p+1}^6(dx^i)^2 + H_5^{NS}(dr^2 + 
r^2d\Omega_2^2)]
\end{equation}
with $H_p = Q_p/r, H_5^{NS} = Q_5^{NS}/r$, where the overall transverse
space is three-dimensional. From the observation that $u$ is proportional
to$\sqrt{r}$ and $b$ is characterized by $\ln r$, whose logarithmic 
behavior is the same as the D5-brane case, we obtain the Penrose limit
metric specified by 
\begin{equation}
\ma = -\frac{1}{4(x^-)^2} [ \sum_{i=1}^{p-1} (x^i)^2 + z^2 +
\left( 1 + \frac{16l^2}{Q_pQ_5^{NS} - l^2} \right) y_1^2 - 
3\sum_{i=p}^6(x^i)^2 ].
\end{equation}
This factor with the characteristic $x^-$-dependence is similar
to that for the D5-brane case, however 
with a slight difference between the
coefficients of $y_1^2$  here and $\sum_{i=1}^2(y^i)^2$ in (\ref{afi}),
which is caused by the different behaviors of $u$ as $u \propto r$ and
$u \propto \sqrt{r}$. 

\section{Conclusion}

We have constructed the Penrose limit metrics 
in closed forms for the near-horizon geometries of
the D1, D3, D5, NS1 and NS5 branes by carrying out explicitly the 
integration in the relation that defines 
the affine parameter $u$ along the
generic null geodesic in terms of the radial 
coordinate $r$, and extracting
its inverse relation with an analytic expression. 
Specially it is observed that 
the radial coordinate is a periodic function of $u$ for the D1 and D3
branes.  The Penrose limit metrics for the D3 and NS5 backgrounds have no
dependence of the light-cone time coordinate $x^-$, while those for
the D1, D5 and NS1 backgrounds have its dependence and show a common 
structrure that they have a singular behavior at $x^-=0$.

We have found that the Penrose limits for the near-horizon geometries
of the marginal 1/4 supersymmetric
bound states consisting of two 
standard intersecting branes, are classified
into three families that are represented by the D1, D3 and D5 types. 
The marginal intersecting system of D$p\perp$D$(4-p)$ with $p=0, 1, 2$
shows the same Penrose limit metric as the D1-brane type, whereas those of
D$p\perp$D$(8-p)(2)$ with $p = 2, 3, 4$ and NS5$\perp$D$p(p-1)$ with 
$p = 1, \cdots 6$  give the Penrose limit 
metrics similar to the D5-brane type.
The other marginal configurations of D$p\perp $D$(6-p)(1)$ with
$p = 1, 2, 3$ and NS1$\parallel$NS5 are so special as to have the same 
Penrose limit metrics of Cahen-Wallach form as the D3-brane type.
The overall transverse spaces of these special configurations are 
four-dimensional so that each harmonic function behaves as $r^{-2}$
with the radial coordinate $r$. Its square $r^{-4}$ for the 
D$p\perp $D$(6-p)(1)$ configuration is associated with the behavior of
harmonic function in the D3-brane configuration.  
Similarly the five-dimensional and three-dimensional overall transverse
spaces of the D$p\perp$D$(4-p)$ and D$p\perp$D$(8-p)(2)$ respectively
yield the $r^{-3}$ and $r^{-1}$ behaviors to each harmonic function, 
whose squares are the behaviors of harmonic functions in the corresponding
D1 and D5 configurations. From these view points, since the backgrounds
of the D2, D4 and D6 branes have the harmonic functions with odd powers
in $r^{-1}$, they are not associated with the intersecting systems of two
D-branes in the Penrose limit. But the standard intersecting system 
of three D2-branes, D2$\perp$D2$\perp$D2 that is a marginal 1/8 
supersymmetric bound state, has three-dimensional overall transverse space
to give a $r^{-1}$ behavior for each harmonic function, whose cube 
indicates the harmonic function for the D4-brane configuration. 
Therefore the Penrose limit metric of D2$\perp$D2$\perp$D2
background is the same as that of the D4 background.
It would be interesting to construct the Penrose limits 
in closed forms on the various non-marginal 
intersecting systems of two branes or the various
non-standard intersecting systems and 
investigate how they are classified.

\end{document}